\documentclass{amsproc}

\theoremstyle{definition}

\theoremstyle{remark}

\numberwithin{equation}{section}

\def\be{\begin{eqnarray}}
\def\ee{\end{eqnarray}}
\def\bee{\begin{eqnarray*}}
\def\eee{\end{eqnarray*}}

\def\cP{{\mathcal P}}
\def\ds{\displaystyle}
 \def\ts{\textstyle}
\def\bra{\langle}
\def\ket{\rangle}

\def\rt2{\ts \frac{1}{\sqrt{2}} }

\def\uparw{\uparrow}
\def\dwnarw{\downarrow}
\def\half{{\textstyle \frac{1}{2}}}
\def\frth{{\textstyle \frac{1}{4}}}

\begin{document}

\title{Pauli Exchange and Quantum Error Correction}

\thanks{\copyright 2000 by author. Reproduction of this article,
  in its entirety,  is permitted for non-commercial
  purposes.}

\author{Mary Beth Ruskai}
\address{Department of
        Mathematics,   University of Massachusetts  Lowell,   Lowell,
        MA  01854 USA} 
\email {bruskai@cs.uml.edu} 
\thanks{Supported in part by National Science
        Foundation Grant DMS-97-06981 and Army Research Office Grant
   DAAG55-98-1-0374}

\date{}


\begin{abstract}
In many physically realistic models of quantum
        computation, Pauli exchange interactions cause a 
special type of two-qubit errors, called exchange errors, 
        to occur as a first order effect of
        couplings within the computer.  We discuss the
physical mechanisms behind exchange errors and codes
designed to explicitly deal with them.
\end{abstract}

\maketitle

\section{Introduction}

Most discussions of quantum error correction assume, at least
implicitly, that errors result from interactions with the
environment\footnote{We consider the  ``environment'' 
as external to the
quantum computer in the sense that  interactions
among qubits, whether or not used in the implementation of
quantum gates, are {\em not} regarded as arising from the
environment.  Many authors follow \cite{KL1} in defining
the environment to include all ``unwanted interactions''.}
and that single qubit errors are 
much more likely than two qubit errors.   
Most discussions also ignore the
Pauli exclusion principle and permutational symmetry of the
states describing multi-qubit systems.   Although this can be
justified by consideration of the full wave function, including
spatial as well as spin components, an analysis (given in \cite{Rusk} )
of these more complete wave functions  suggests 
that more attention
should be given to the  effect of exchange interactions
within a quantum computer.   Interactions between identical
particles can cause an error in two bits simultaneously as
a first order effect.
Moreover, because they result from interactions within the quantum
        computer, exchange errors cannot be reduced by
        better isolating the quantum computer from its
        environment.
The effect of a (single) exchange error is to flip two bits
if, and only if, they are different.   It is a non-classical
type of error in the sense that it  arises directly from a physical
mechanism which occurs only in the case of identical particles
which follow the rules of quantum theory.    
If classical systems were to exhibit this type of behavior,
they would require unusual correlations which do not normally
occur from first-order couplings.

Schemes for fault-tolerant computation \cite{Gott2,Pk1}
have been developed which treat two-bit, and even multi-bit, errors.  
(See, e.g., \cite{Kt,KL1,KLV,St} and references
in \cite{Gott3,Pk2} and at the end of Chapter 10 of \cite{NC}.)
However, many
of these, such as those arising from concatenated codes
\cite{AB,KL2,Pk2} require 
a large number of physical bits to represent one logical bit.
Steane, in particular, has emphasized the problems
associated with the size of repeatedly concatenated codes
and proposed new methods \cite{St} for coding $m$ logical bits in $n$
qubits.
Furthermore, threshold estimates \cite{AB,Gott2,KLZ,Pk1} 
are generally based on the
assumption that two-bit errors are second order effects
resulting from uncorrelated interactions with the 
environment.
In those situations where exchange errors are important 
shorter codes that explicitly address exchange errors can
be effective.

A very different approach to fault tolerant computation is based
on the assumption of highly correlated errors at low temperature,
allowing the use of ```decoherence free subspaces'' (DFS)
\cite{DFS1,DFS2}.  Shortly after \cite{Rusk} was posted,
Lidar, et al observed \cite{DFS3} that the existing schemes for
concatenating a DFS code with a standard 5-qubit code for correcting
single-bit errors \cite{DFS2} could also correct exchange errors.  This is
because exchange errors on physical qubits appear as single Pauli errors
on the logical bits used in DFS codes.  Subsequently, they \cite{DFS4}
turned this idea around to show how exchange interactions could be
used to construct universal gates within the DFS scheme for quantum
computation.

In section \ref{sect:full}  we first review some of the 
basic principles 
underlying permutational symmetry of multi-particle
quantum wave functions and then show how this leads to
exchange errors.   In Section \ref{sect:error} we discuss
the issues associated with correction of exchange errors
and present an explicit (non-additive) 9-bit code which can correct both
exchange errors and all one-qubit errors.  Section \ref{sect:newcode}
contains an ambitious proposal for constructing
powerful new codes using irreducible representations
of the symmetric group.

\section{The full  wave function }\label{sect:full}

\subsection{Permutational symmetry}

A (pure) state of a quantum mechanical particle with spin $q$
corresponds to a one-dimensional subspace of the Hilbert space
${\mathcal H} = {\bf C}^{2q+1} \otimes L^2({\bf R}^3)$ and is typically 
represented by a vector in that subspace.   The state of a
system of $N$ such particles is then represented by a vector
$\Psi(x_1,x_2,\ldots,x_N)$ in ${\mathcal H}^N$.  However, when dealing
with identical particles $\Psi$ must also satisfy the Pauli
principle, i.e., it must be symmetric or anti-symmetric under
exchange of the coordinates $x_j \leftrightarrow x_k$ so that, e.g., 
\begin{eqnarray}
  \Psi(x_2,x_1,\ldots,x_N) = \pm \Psi(x_1,x_2,\ldots,x_N) .
\end{eqnarray}
depending
on whether the particles in question are bosons (e.g. photons)
or fermions (e.g., electrons).  In either case, we can write
the full wave function in the form
\begin{eqnarray} \label{eq:fullspin}
\Psi(x_1,x_2,\ldots,x_N) 
= \sum_k   
   \chi_k(s_1,s_2,\ldots,s_N) \Phi_k({\bf r}_1,{\bf r}_2,\ldots,{\bf r}_N)
\end{eqnarray}
where the ``space functions" $\Phi_k$ are elements of 
$L^2({\bf R}^{3N})$, the ``spin functions" $\chi_k$ are
\footnote{A spin state $\chi$  looks formally like
 a (possibly entangled) N-qubit state.  However, unlike qubits
which involve an implicit spatial component, we want only vectors
in $[{\bf C}^{2q+1}]^N$ itself.}  in
 $[{\bf C}^{2q+1}]^N$ and $x_k = ({\bf r}_k,s_k)$
with ${\bf r}$ with a vector  in ${\bf R}^3$ and 
  $s_k$  (called the spin coordinate) an element of 
$\{ 0, 1, \ldots 2q \} $corresponding to spin values 
going from $-\half q$ to  $+\half q$ in integer steps.  It
is not necessary that  $\chi$ and $\Phi$ each satisfy the Pauli 
principle; indeed, when $q= \half$ so that $2q+1 = 2$ and we
are dealing with ${\bf C}^2$ it is {\em not} possible for $\chi$
to be anti-symmetric when $N \geq 3$.   Instead, we 
expect that  $\chi$ and $\Phi$  satisfy certain duality conditions
which guarantee that $\Psi$ has the correct permutational symmetry.
In the case of anti-symmetric functions there is an extensive
literature about functions in which the $\chi_k$ and
$\Phi_k$ are bases for irreducible representations of $S_n$
with dual Young tableaux.

With this background, we now restrict attention to the
important special case in which $q = \half$ yielding 
two spin states labeled\footnote{These labels are the reverse of the usual
physicists's convention; in essence, the  convention in
quantum computation is to label the eigenvectors of $\sigma_z$
so that the $\hbox{eigenvalue} = e^{i \, \hbox{label}}$.}
 so that $s = + \half$ corresponds to $|0\ket$  
and  $s = - \half$ corresponds to $|1\ket$, 
 and the particles are electrons so that $\Psi$ must
be anti-symmetric.   


To emphasize the distinction between a pure spin state as an 
element of ${\bf C}^2$ and a spin associated with a 
particular qubit or spatial wave function, we will
replace $|0\ket$ and $|1\ket$ by $\uparw$ and $\dwnarw$ respectively.
The notation $|01\ket$ then describes a two-qubit state in which the
particle in the first qubit has spin ``up'' ($\uparrow$) and that
in the second has spin ``down'' ($\downarrow$).  What does it
mean for a particle to ``be'' in a qubit?   A reasonable answer
is that each qubit is identified by the spatial  component of its
wave function  $f_A({\bf r})$ where $A, B, C \ldots$ label the
qubits and wave functions for different qubits are orthogonal.  
Thus, if the qubits did not correspond to identical particles
we would have 
$ |01\ket = f_A({\bf r_1})\!\uparw \, f_B({\bf r_2})\!\dwnarw$.
In the more realistic situation of identical particles
\begin{eqnarray}\label{psi01}
  |01\ket = \rt2 \big( f_A({\bf r_1})\!\uparw \, f_B({\bf r_2})\!\dwnarw
   \pm  \, f_B({\bf r_1})\!\dwnarw \, f_A({\bf r_2})\!\uparw  \big).
\end{eqnarray} 
with the plus  sign ($+$) for bosons and the minus sign ($-$)
for fermions.  We will henceforth consider the special case of
electrons, which are fermions, in which case the  antisymmetric 
function given by (\ref{psi01}) is called a Slater determinant.
Note that a function of the form (\ref{psi01}) has the important
property that the electron whose spatial function is
$f_A$ always has spin ``up'' regardless of whether its coordinates are
labeled by $1$ or $2$.   Although (\ref{psi01}) is not a simple product,
but a special type of superposition which
is the anti-symmetrization of a product,  it behaves in some ways like
a product state.  It should be contrasted with the a true entangled
Bell state such as
\begin{eqnarray} \label{eq:bell}
\rt2 \big[ |01\ket - |10\ket   \big]   
 & = &  \half \big( 
    f_A({\bf r_1})\!\uparw \, f_B({\bf r_2})\!\dwnarw
   -   f_B({\bf r_1})\!\dwnarw \, f_A({\bf r_2})\!\uparw  \\
 & ~ & ~~~\,\,\,\,  - f_A({\bf r_1})\!\dwnarw \, f_B({\bf r_2})\!\uparw
   + f_B({\bf r_1})\!\uparw \, f_A({\bf r_2})\!\dwnarw  \big)  \nonumber
\end{eqnarray} 
which is a superposition of two Slater determinants or four
products.  

It may be useful to observe that  (\ref{eq:bell}) has the 
form of a wave
function associated with an entangled state shared by ``Alice''
and ``Bob'' when $f_A$ describes a particle localized near
Alice and $f_B$ a particle localized near Bob, and
discuss its interpretation in that situation.  When Alice uses
 a detector in her lab to measure the spin, she also implicitly
makes a measurement of the spatial function, i.e., a measurement
which projects onto spatial functions localized in her lab.
   She may get
electron \#1 with spin ``up'' with probability $\frth$ or
electron \#2 with spin ``up'' with probability $\frth$.  However,
there is no physical way to distinguish these two possibilities.
The net result is a measurement of spin ``up'' in Alice's lab
with total probability $\half$.  The other two states in the
superposition would correspond to measuring some electron in her
lab with spin ``down'', also with net probability $\half$.
Once Alice has made a measurement, a corresponding measurement by Bob 
always yields the opposite spin.

Returning to (\ref{psi01}), we note that it can be
 rewritten in  the form (\ref{eq:twobit}) as 
\begin{eqnarray}\label{eq:twobit}
|01\ket =  {\textstyle \rt2 } 
  [ \chi^+\!(s_1,s_2) \phi^-\!({\bf r_1},{\bf r_2}) +
 \chi^-\!(s_1,s_2) \phi^+ \!({\bf r_1},{\bf r_2}) ]
\end{eqnarray} 
where  $\chi^{\pm}= \rt2 \left[ \uparw \dwnarw \pm \dwnarw \uparw \right]$
denote the indicated Bell-like spin states and 
\begin{eqnarray*}
\phi^{\pm} = \rt2 \left[f_A({\bf r_1}) f_B({\bf
r_2}) \pm  f_B({\bf r_1}) f_A({\bf r_2}) \right] .
\end{eqnarray*}
It should be emphasized that the reduction to a simple expression 
of the form (\ref{eq:twobit}), in which each term in the product 
is either an {\em antisymmetric} spin function times a {\em symmetric}
 spatial functions or vice-versa, is possible only when $N = 2$.  For more than
two electrons, more complex expressions, of the form 
(\ref{eq:fullspin}) are needed.

\subsection{The origin of Pauli exchange errors}

We now describe the origin of Pauli exchange
errors by analyzing the two-qubit case in detail, under
the additional simplifying assumption that the Hamiltonian
is spin-free.  This is certainly not realistic; quantum
computers based upon spin will involve magnetic fields
and hence, not be spin-free.   However,  the assumption of 
a spin-free  Hamiltonian $H$, merely implies that
the time development of  (\ref{psi01}) is determined 
 by $e^{-iHt} \phi^{\pm}$, and this
 suffices to illustrate the principles involved.
With a spin-dependent Hamiltonian  the time development  
$e^{-iHt} \chi^{\pm}$ would also be non-trivial.

 We will also assume that the qubits are formed 
using charged particles, such
as electrons or protons, so that $H$ includes a term corresponding
to the $\frac{1}{r_{12}} \equiv \frac{1}{|{\bf r_1}-{\bf r_2}|}$
long-range Coulomb interaction. The Hamiltonian will be
symmetric so that the states
$\phi^{\pm}$ retain their permutational symmetry; however, the
 interaction term implies that they
 will not retain the simple form of symmetrized (or anti-symmetrized)
product states.  Hence, after some time the states $\phi^{\pm}$
 evolve into
\begin{subequations} 
 \begin{eqnarray} \label{eq:phi.time} 
 \Phi^- ({\bf r}_1,{\bf r}_2 ) & = & \sum_{m < n} c_{mn} \rt2 \left[
  f_m({\bf r_1}) f_n({\bf r_2}) -  f_n({\bf r}_1) f_m({\bf r}_2) \right] \\
\Phi^+ ({\bf r}_1,{\bf r}_2 ) & = & \sum_{m \leq n} d_{mn} \rt2 \left[
 f_m({\bf r_1}) f_n({\bf r_2}) +  f_n({\bf r}_1) f_m({\bf r}_2) \right].
\end{eqnarray}
\end{subequations} 
 where $f_m$ denotes any orthonormal basis whose first two elements
are $f_A$ and $f_B$ respectively.
There is no reason to expect that $c_{mn} = d_{mn}$ in general.  
On the contrary, only the symmetric sum includes pairs with  $m = n$.
Hence if one $d_{mm} \neq 0$, then one must have some 
$c_{mn} \neq d_{mn}.$
Inserting (\ref{eq:phi.time}) in (\ref{eq:twobit}) yields 
\begin{eqnarray}
 e^{-iHt} |01 \ket  & =  &
  \frac{c_{AB} + d_{AB}}{2}
  \big( f_A({\bf r_1}) \! \uparw  f_B({\bf r_2}) \! \dwnarw 
  -  f_B({\bf r_1})\! \dwnarw   f_A({\bf r_2})\! \uparw \big) 
    \nonumber  \\
  & ~ &  + \, \frac{c_{AB} - d_{AB}}{2}       
  \big( f_B({\bf r_1})\! \uparw  f_A({\bf r_2})\! \dwnarw 
  -  f_A({\bf r_1})\! \dwnarw  f_B({\bf r_2})\! \uparw  \big) \nonumber 
   + \Psi^{\rm Remain}\\
 & = & \frac{c_{AB} + d_{AB}}{2}|01 \ket + \frac{c_{AB} - d_{AB}}{2} |10 \ket 
   + \Psi^{\rm Remain} 
\end{eqnarray} 
where $\Psi^{\rm Remain}$ is orthogonal to $\phi^{\pm}$.

A measurement of qubit-A  corresponds to projecting onto  $f_A$.  
Hence a measurement of qubit-A on the state (\ref{eq:twobit})
yields spin ``up'' with probability ${\frth |c_{AB} + d_{AB}|^2}$
and spin ``down'' with probability  $\frth|c_{AB} - d_{AB}|^2$.   Note
 that the {\em full} wave function is {\em necessarily} an {\em entangled}
state and that the measurement process leaves the system in
state $|10 \ket$ or $|01 \ket$ with probabilities 
 $\frth|c_{AB} \pm d_{AB}|^2$ respectively, i.e., a subsequent measurement
of qubit-B always yields the opposite spin.  With probability
 $\frth|c_{AB} - d_{AB}|^2$ the initial state $|10 \ket$ has been
converted to $|01 \ket$. 

Although the probability of this may be small, it is {\em not} zero.  
Moreover, it would seem that any
implementation which provides a mechanism for two-qubit
gates would not allow the qubits to be so isolated as to
preclude interactions between particles in different 
qubits\footnote{Although the gates themselves require interactions, 
we expect these to be short-lived and well-controlled, i.e.,  in a
well-designed quantum computer the gates themselves should
not be a significant source of error.  However, the process
of turning gates on and off could induce errors in other
qubits.  We do not consider this error mechanism.}.
In general, one would expect qubits to be less isolated from
each other than from the external environment 
so that the interaction between a single pair of qubits
would be greater than between a qubit and a particle in the
environment.  However, the environment consists of a huge
number of particles (in theory, the rest of the world) and 
it may well happen that the number of  environmental
particles which interact with a given qubit is several orders
of magnitude greater than the number of qubits,  giving a
net qubit-environment interaction which is greater than a
typical qubit-qubit interaction.    On the other hand, the
number of qubit-qubit interactions grows quadratically with
the size of the computer.  Thus, prototype quantum computers,
using only a few qubits, may not undergo exchange errors at
the same level as the larger computers needed for real computations.

It is worth emphasizing that when
 the implementation involves charged particles,
whether electrons or nuclei, the interaction always includes a
contribution from the $\frac{1}{r_{12}}$ Coulomb potential
which is known to have long-range\footnote{This is
because even when $f$ and $g$ have non-overlapping
compact support $[a,b]$ and $[c,d]$ respectively,
such expectations as
$\int\int |f({\bf r_1})|^2 \frac{1}{|{\bf r_1}-{\bf r_2}|} f({\bf r_2})|^2 $ 
will be non-zero
because the integrand is non-zero on $[a,b] \times [c,d]$.
Non-overlapping initial states will not prevent the system from
evolving in time to one whose states are {\em not} simple products or
(in the case of fermions) Slater determinants! }  effects.
 Screening may
reduce the effective charge, but it will not, in general,
remove the basic long-range behavior of the Coulomb
interaction. 

Precise estimates of exchange errors require  more detailed models
of the specific experimental implementations.   The role of
long-range Coulomb effects (for which exchange errors
grow quadratically with the size of the computer) suggests
that implementations involving neutral particles may be
advantageous for minimizing exchange errors.  This would
include both computers based on polarized photons (rather than
charged particles) and more innovative schemes,  such as
Briegel, et al's  proposal \cite{BCJCZ}  using optical lattices.
On the other hand, the ease with which exchange errors can be
corrected using appropriate 9-qubit codes, suggests that dealing
with exchange interactions need not be a serious obstacle.

\section{Correcting Exchange Errors} \label{sect:error}

A Pauli exchange error is a special type of ``two-qubit" error
 which has the same effect as ``bit flips" if (and {\it only} if)
they are different.  Exchange of bits  $j$ and $k$ is equivalent
to acting on a state with the operator
\begin{eqnarray}\label{def:Ejk}
  E_{jk} = \half \Big( I_j \otimes I_k + Z_j \otimes Z_k
   + X_j \otimes X_k  + Y_j \otimes Y_k \Big)
\end{eqnarray}
where $X_j, Y_j, Z_j$ denote the action of the Pauli matrices
$\sigma_x,  \sigma_y, \sigma_z$ respectively on the bit $j$.

\subsection{Example: the 9-bit Shor code}

As an example of potential difficulties with existing codes, 
 consider   the simple 9-bit code of  Shor \cite{Shor}
   \begin{subequations} 
\begin{eqnarray}  \label{Shorcode} 
| c_0 \ket    & = & |{\bf 000} \ket  + |{\bf 011} \ket + 
   |{\bf 101} \ket  + |{\bf 110} \ket  \\
 | c_1 \ket & = &    |{\bf 111} \ket  +
    |{\bf 100} \ket + |{\bf 010}\ket  + |{\bf 001} \ket 
\end{eqnarray}
\end{subequations}
where boldface denotes a triplet of 0's or 1's.
It is clear that these code words are invariant under exchange
of electrons within the 3-qubit triples (1,2,3),  (4,5,6), or (7,8,9).
To see what happens when electrons in different triplets are
exchanged, consider the exchange $E_{34}$ acting on $| c_0 \ket $.
This yields $ |000\,000\,000\ket + |001\,011\,111\ket +
|110\,100\,111\ket  +  |111\,111\,000\ket $ so that
   \begin{subequations} 
\begin{eqnarray} 
  E_{34}  | c_0  \ket & = &  | c_0 \ket +  Z_2  | c_0 \ket +
    |001\,011\,111\ket + |110\,100\,111\ket   \\
      E_{34} | c_1 \ket  & = &  | c_1 \ket -  Z_2  | c_1 \ket 
    + |110\,100\,000\ket + |001\,011\,000 \ket  
\end{eqnarray}
\end{subequations}
If $|\psi \ket = a | c_0 \ket + b | c_1 \ket$ is a superposition
of code words, 
\begin{eqnarray*} 
   E_{34} |\psi \ket  = \half \Big( |\psi \ket +  
     Z_8 | \tilde{\psi} \ket  \Big) + \rt2 | \gamma \ket
\end{eqnarray*}
where $ | \tilde{\psi} \ket = a | c_0 \ket - b | c_1 \ket$ differs
from $\psi$ by a ``phase error" on the code words and  $| \gamma \ket$
is orthogonal to the space of codewords and single bit errors.
Thus, this code cannot reliably distinguish between an exchange
error $E_{34}$ and a phase error on any of the last 3 bits.  This
problem occurs because if one tries to write
$ E_{34} | c_0 \ket = \alpha  | c_0 \ket + \beta | d_0 \ket $
with $| d_0 \ket $  orthogonal to $| c_0 \ket$, then  one can
not also require that $| d_0 \ket $   be orthogonal to $| c_1 \ket$.

\subsection{Conditions for error correction}

Before discussing specific codes  for correcting exchange errors, we first
review some of the basic principles of error correction. 
In order to be able to correct a given class of errors, we identify
a set of basic errors $\{ e_p \}$  in terms of which all other errors can
be written as linear combinations.  In the case of unitary transformations
on single bit, or one-qubit errors, this set usually consists of
$X_k, Y_k, Z_k ~~ (k= 1 \ldots n)$ where $n$ is the number of qubits in
the code and $X_k, Y_k, Z_k$ now denote 
$I \otimes I \otimes I \ldots\otimes  \sigma_p \otimes \ldots \otimes I$
where $ \sigma_p $ denotes one of the three Pauli matrices acting
on qubit-k.  If we let $e_0 = I$ denote the identity, then a
sufficient condition for error correction is
\begin{eqnarray}\label{err.orthog}
\bra e_p C_i | e_q C_j \ket = \delta_{ij} \delta_{pq}
\end{eqnarray}
However, (\ref{err.orthog})  can be replaced 
\cite{CDSW,CRSS,KL1} by the weaker condition
\begin{eqnarray}\label{err.suff}
  \bra e_p C_i | e_q C_j \ket = \delta_{ij} d_{pq} .
\end{eqnarray}
where the matrix $D$ with elements $d_{pq}$ is independent of $i,j$.
When considering Pauli exchange errors, it is natural to seek codes
which are invariant under some subset of permutations.  This is
clearly incompatible with (\ref{err.orthog}) since some of the exchange
errors will then satisfy $ E_{jk} | C_i \ket = | C_i \ket $.
Hence we will need to use (\ref{err.suff}).

The most common code words have the
property that $| C_1 \ket $ can be obtained from $| C_0 \ket$
by exchanging all 0's and 1's. For such codes, it is not
hard to see that
 $ \bra C_1 | Z_k C_1 \ket =  - \bra C_0 | Z_k C_0 \ket $
which is consistent with (\ref{err.suff}) if and only if 
it is identically zero.
Hence even when using (\ref{err.suff}) rather than (\ref{err.orthog})
it is necessary to require 
\begin{eqnarray}\label{eq:dualphase}
  \bra C_1 | Z_k C_1 \ket =  - \bra C_0 | Z_k C_0 \ket = 0
\end{eqnarray}
when the code words have this type of $0 \leftrightarrow 1$ duality.

If  the basic error set has size
$N$ (i.e., $p = 0, 1 \ldots N-1$), then a two-word code requires
codes which lie in a space of dimension at least $2N$.  For 
the familiar case of single-bit errors $N = 3n+1$ and, since an
n-bit code word lies in a space of dimension $2^n$, any code must
satisfy $3n+1 < 2^{n-1}$ or $n \geq 5$.
There are $n(n-1)/2$ possible single exchange
errors compared to $9n(n-1)/2$ two-bit errors of all types.  
Thus, similar dimension arguments would imply that codes
which can correct all one- and two-bit errors must satisfy
 $2N = 9n(n-1) + 2(3n+1) \leq 2^n$ or $n \geq 10$.
The shortest code  known \cite{CRSS}
which can do this has n = 11.  
We will see that, not surprisingly, correcting both one-bit and Pauli exchange
errors,  can be done with shorter codes than required to correct all 
two-bit errors.  

However, the dimensional analysis above need not yield the best
bounds when exchange errors are involved.  Consider
 the simple code $ | C_0 \ket =  |000\ket, | C_1 \ket =  |111\ket $
which is optimal for single bit flips (but
can not correct phase errors).  In this case $N = n+1$ and $n = 3$
yields equality in  $2(n + 1) \leq 2^n$.
But, since this code is invariant
under permutations, the basic error set can be
expanded to include all 6 exchange errors $E_{jk}$ for a total
of $N = 10$ without increasing the length of the code words.

\subsection{Permutationally invariant  codes}\label{sect:code9}

We now  present  a 9-bit code code which  
can handle both  Pauli exchange errors and all one-bit errors.
It is based on the realization that codes
which are invariant under permutations are impervious to
Pauli exchange errors. 
Let 
\begin{subequations}  \label{code9}
 \begin{eqnarray}
| C_0 \ket & = &  |000\, 000\, 000\ket + 
    \frac{1}{\sqrt{28}} \sum_{\cP} |111\,111\,000\ket \\
 | C_1 \ket & = &  |111\,111\,111\ket + 
     \frac{1}{\sqrt{28}} \sum_{\cP} |000\,000\,111\ket 
\end{eqnarray} 
\end{subequations}
where $\sum_{\cP}$ denotes the sum over all permutations of the
indicated sequence of 0's and 1's and it is understood that we
count permutations which result in identical vectors only once.
This differs from the  9-bit Shor code  in that 
{\em all} permutations of $|111\,111\,000 \ket$ are included,
rather than only three.  The normalization of the code words is
$$\bra  C_i |  C_i \ket  = 1 + \frac{1}{28} \binom{9}{3} = 4 \, .$$

The coefficient $1/\sqrt{28}$ is needed to satisfy (\ref{eq:dualphase}).
Simple combinatorics implies
\begin{eqnarray*}
\bra  C_i | Z_k C_i \ket = 
  (-1)^i \left[ 1 - \frac{1}{3}\binom{9}{3}\frac{1}{28} \right]
  = 0.
\end{eqnarray*}
Moreover,
\begin{eqnarray}\label{eq:phase.mat}
\bra  Z_k C_i | Z_{\ell} C_i \ket = 
  1 + \delta_{k \ell} \binom {9}{3}\frac{1}{28}  = 1 + 3 \delta_{k \ell} .
\end{eqnarray}
The  second term in (\ref{eq:phase.mat}) is zero 
when $k \neq \ell$ because of
the fortuitous  fact that there are exactly the same
number of positive and negative terms.  If, instead, we had
used all permutations of $\kappa$ 1's in $n$ qubits, this term would be
 $\ds{\frac{(n-2 \kappa)^2 - n}{n(n-1)} }\binom{n}{\kappa}$ when
$k \neq \ell$.
 
Since all components of $|C_0\ket$ have $0$ or  $6$ bits equal to 1, 
any single bit flip acting on $|C_0\ket$, will yield a vector
whose components have $1,  5$, or $7$ bits equal to 1 and is
thus orthogonal to $|C_0\ket$, to $|C_1\ket$, to a bit flip acting
on $|C_1\ket$ and to a
phase error on either  $|C_0\ket$ or $|C_1\ket$.
Similarly, a single bit flip on $|C_1\ket$ will yield a vector
orthogonal to $|C_0\ket$, to $|C_1\ket$, to a bit flip acting
on $|C_0\ket$ and to a  phase error on $|C_0\ket$ or $|C_1\ket$. 
This suffices to ensure that (\ref{err.orthog}), and hence
(\ref{err.holds}), holds if $e_p$ is $I$ or some $Z_k$ and
 $e_q$ is one of the $X_{\ell}$ or $Y_{\ell}$.

 However, single bit flips on a given code word
need  not be mutually orthogonal.
To find $\bra  X_k C_i | X_{\ell} C_i \ket$ when $k \neq \ell$,
consider
\begin{eqnarray}\label{eq:bit.mat}
\bra  X_k \, (\nu_1 \nu_2 \ldots \nu_9) \, | \,
                 X_{\ell} \, (\mu_1 \mu_2 \ldots \mu_9) \ket. 
\end{eqnarray}
where $\nu_i, \mu_i$ are in $0,1$. 
This will be nonzero only when 
$\nu_k = \mu_{\ell} = 0, ~~  \nu_{\ell} =  \mu_k = 1$ or
$\nu_k = \mu_{\ell} = 1, ~~  \nu_{\ell} =  \mu_k = 0$
and the other $n-2$ bits are equal.  From $\sum_{\cP}$ with 
$\kappa$ of $n$ bits
equal to 1, there are $2 \binom{n-2}{\kappa-1}$ such terms.  Thus, for the
code (\ref{code9}), there are 42 such terms which yields an
inner product of $\frac{42}{28} = \frac{3}{2}$ when $k \neq \ell$.
We similarly find that
$$\bra  Y_k C_i | X_{\ell} C_i \ket 
  = - i \bra  X_k Z_k C_i | X_{\ell} C_i \ket = 0 \,\,\,\,
\hbox{for all} ~ k \neq \ell$$ 
because exactly half of the
terms analogous to (\ref{eq:bit.mat}) will occur with a positive
sign and half with a negative sign, yielding  a net  inner product
of zero.
We also find 
$$\bra  Y_k C_i | X_k C_i \ket = - i \bra  X_k Z_k C_i | X_k C_i \ket
  = -i \bra  Z_k C_i |  C_i \ket = 0 $$
so that 
$$\bra  Y_k C_i | X_{\ell} C_i \ket = 0 \,\,\,\,
\hbox{for all} ~ k, \ell.$$

These  results imply that  (\ref{err.suff}) holds and 
that the matrix $D$  is block diagonal with the form
\begin{eqnarray}
D = \left( \begin{array}{cccc}  
   D_0 & 0  & 0 & 0 \\
   0 &  D_X & 0 & 0 \\
   0 & 0 & D_Y & 0  \\
   0 & 0 & 0 & D_Z \end{array} \right)
\end{eqnarray}
where $D_0$ is the $37 \times 37$ matrix corresponding to the 
identity and the 36 exchange errors, and
$D_X, D_Y, D_Z$ are $9 \times 9$ matrices corresponding
respectively to the $X_k, Y_k, Z_k$ single bit errors.
One easily finds that $d^0_{pq} = 4$ for all $p,q$ so that
$D_0$ is is a multiple of a one-dimensional projection.
The $9 \times 9$ matrices $D_X, D_Y, D_Z$ all have 
$d_{kk} = 4$ while for $ k \neq \ell$, $ d_{k \ell} =  3/2$ 
in $D_X$ and $D_Y$ but $d_{k \ell}  = 1$ in $D_Z$.
Orthogonalization of this matrix is straightforward.
Since $D$ has rank $28 = 3 \cdot 9 + 1$, we are using only
a $54 < 2^6$ dimensional subspace of our $2^{9}$ dimension space.

The simplicity of codes which 
 are invariant under permutations makes them attractive.
However, there are few such codes.   All code words must have the form
\begin{eqnarray}\label{eq:permcode}
  \sum_{\kappa=0}^n a_\kappa \sum_{\cP} 
   |\underbrace{ 1  \ldots 1}_\kappa \underbrace{ 0 \ldots 0}_{n-\kappa} \ket.
\end{eqnarray}
 Condition 
(\ref{err.suff})  places some severe restrictions
on the coefficient $a_\kappa.$  For example, in (\ref{code9}) 
only $a_0$ and $a_6$
are non-zero in $|C_0\ket$ and only $a_3$ and $a_9$ in $|C_1\ket$.
If we try to change this so that $a_0$ and $a_3$ are non-zero
in $|C_0\ket$, i.e., 
\begin{subequations}  \label{code9var}
 \begin{eqnarray} 
| C_0 \ket & = &  a_0 |000\, 000\, 000\ket + 
     a_3  \sum_{\cP} |111\,000\,000\ket \\
 | C_1 \ket & = & a_9 |111\,111\,111\ket + 
     a_6 \sum_{\cP} |000\,111\,111\ket 
\end{eqnarray}  \end{subequations}  
then it is {\em not}
possible to satisfy (\ref{eq:dualphase}).

The 5-bit error correction code in \cite{BDSW,LMPZ} does not have the 
permutationally invariant form
(\ref{eq:permcode}) because the code words include components
of the form $\sum_{\cP} \pm | 11000 \ket $, i.e., not all terms
in the sum have the same sign.
 The non-additive 5-bit error {\em detection} code in  \cite{RHSS}  
also requires  changes in the $\sum_{\cP} \pm | 10000 \ket$ term.
Since such sign changes seem needed to satisfy (\ref{eq:dualphase}),
one would not expect that 5-bit codes can  handle 
Pauli exchange errors.  In fact, Rains \cite{Rains} has shown that
the 5-bit error correction code is essentially unique, which implies
that no 5-bit code can correct both all 1-bit errors and exchange
errors.  In \cite{Rusk} the possibility of  7-bit codes of the form
(\ref{code9var}) was raised.
However, Wallach \cite{W} has obtained convincing evidence that
no permutationally invariant 7-bit code can correct all one-qubit errors.


\subsection{Proposal for a new class of codes}\label{sect:newcode}

Permutational invariance, which is based on
a one-dimensional representation of the symmetric group, 
is not the only approach to exchange errors.  Our analysis of 
(\ref{Shorcode}) suggests a construction which we
first describe in over-simplified form.   Let 
$|c_0 \ket, |d_0 \ket, |c_1 \ket, |d_1 \ket$  be four  mutually orthogonal
n-bit vectors such that $|c_0 \ket,  |c_1 \ket$ form a  code
for one-bit errors and $|c_0 \ket, |d_0 \ket$ and $|c_1 \ket, |d_1 \ket$
are each bases of a two-dimensional representation of the symmetric
group $S_n$.  If $|d_0 \ket$ and $|d_1 \ket$  are also
orthogonal to one-bit errors on the code words, then this code can correct
Pauli exchange errors as well as one-bit errors.  If, in addition,
the vectors $|d_0 \ket,  |d_1 \ket$ also form a code
isomorphic to $|c_0 \ket,  |c_1 \ket$ in the sense that the matrix
$D$ in (\ref{err.suff}) is identical for both codes, then the code
should also be able to correct products of one-bit and 
Pauli exchange errors.

However, applying this scheme to an n-bit code requires a
non-trivial irreducible representation of $S_n$ of which
the smallest has dimension $n-1$.
Thus we will seek a set of $2(n-1)$ mutually orthogonal vectors denoted
$|C_0^m \ket,  |C_1^m \ket ~(m= 1 \ldots n-1)$ such that
$|C_0^1 \ket,  |C_1^1 \ket$ form a code for one bit errors and
$|C_0^m \ket ~(m= 1 \ldots n-1)$ and $|C_1^m \ket ~(m= 1 \ldots n-1)$ 
each form basis of the same irreducible representation of $S_n$.
Such code will be able to correct {\em all} errors which permute
qubits; not just single exchanges.
If, in addition, (\ref{err.suff}) is extended to 
\begin{eqnarray}\label{err.suff.ext}
  \bra e_p C_i^m | e_q C_j^{m'} \ket = \delta_{ij} \delta_{mm'} d_{pq}
\end{eqnarray}
with the matrix $D = \{D_{pq}\}$ independent of both $i$ and $m$,
then this code will also  be able to correct 
 products of one bit errors and permutation errors.

In the construction proposed above, correction of exchange and one-bit errors
would require a space of dimension $2(n-1)(3n+1) \leq 2^n$ or
$n \geq 9$.  If codes satisfying (\ref{err.suff.ext}) exist, 
 they could correct {\em all} permutation errors
as well as products of permutations and one-bit errors (which includes a
very special subclass of 3-bit errors and even a few higher ones).   
Thus exploiting permutational symmetry may yield powerful new codes.

In some sense, the strategy proposed here is 
the opposite of that of Section \ref{sect:code9} (despite the fact that both are
 based on representations of $S_n$).
In  Section \ref{sect:code9} we sought code words with the 
maximum symmetry of being invariant under all permutatations.
Now, we seek instead, a pair of dual code words  
$|C_0^1 \ket,  |C_1^1 \ket $ with ``minimal'' symmetry
in the sense that a set of generators of $S_n$ acting on each of these
code words yields an orthogonal basis for a non-trivial irreducible representation 
of $S_n$.   If the code words $|C_0^m \ket,  |C_1^m \ket ~~ n=2\ldots n-1$
can be obtained in this way, then each pair should also be a
single-bit error correction code, as desired.

\subsection{Non-additive codes}

Most existing codes used for quantum error correction are obtained
by a process \cite{CRSS,Gott1,Gott3} 
through which the codes can be described in terms
of a subgroup, called the {\em stabilizer}, of the error group.
Such codes are called ``stabilizer codes'' or ``additive codes''.
In \cite{RHSS} an example of a non-additive 5-bit code was given,
establishing the existence of non-additive codes.  However, this
was only an error detection code and, hence, less powerful than
the 5-bit error correction code \cite{BDSW,LMPZ} obtained using the stabilizer
formalism.  Subsequently, V. P. Roychowdhury and F.  Vatan \cite{RV}
showed that many non-additive codes exist; however, it was not clear
how useful such codes might be.

H. Pollatsek \cite{P} has pointed out that the 9-qubit code 
(\ref{code9}) is a  non-additive code.   This establishes that 
non-additive codes may well have an important role to play in
quantum error correction, particularly in situations in which
exchange errors and permutational symmetry are important.

Her argument is based on the observation that the set of vectors
which occur in each $\sum_{\cP}$ in (\ref{code9}) spans the vector
space of binary 9-tuples ${\bf Z}_2^9$.  More generally let
$\Gamma_{\kappa,n}$ denote the set of all vectors 
${\bf a} = (a_1, a_2, \ldots a_n)$ in ${\bf Z}_2^n$
with precisely $\kappa$ of the $a_j$ taking the
value $1$ and $n - \kappa$ the value $0$ as in (\ref{eq:permcode}).
Then span$\{ \Gamma_{\kappa,n} \} = {\bf Z}_2^n$ if $\kappa \neq 0,n$.
By definition, 
 an additive (or stabilizer) code forms an eigenspace for an abelian
subgroup 
$S$ of the error group 
$E = \{ i^{\ell} X({\bf a})Z({\bf b}) : {\bf a}, {\bf b} \in {\bf Z}_2^9 \}$.
When the stabilizer  $S$ consists of only the scalar multiples
of the identity $I$,
then the corresponding eigenspace is all of ${\bf C}^{2^9}$.
  Consequently, to show that  (\ref{code9})
is {\em not} a stabilizer code, it suffices to show that no vector 
of the subspace spanned by the codewords $|C_0 \ket$ and $|C_1 \ket$
can be an eigenvector for an element of $E$ other than $I$.

 The image of $|C_0 \ket$ under $X(a)Z(b)$ is
\begin{eqnarray} 
 X({\bf a})Z({\bf b}) |C_0 \ket = 
| {\bf a} \ket + \frac{1}{\sqrt{28}}\sum_{ {\bf v} \in 
\Gamma_{6,9}} (-1)^{{\bf b} \cdot {\bf v}} | \, {\bf v} + {\bf b} \, \ket.
 \end{eqnarray}
If $ X({\bf a})Z({\bf b})|C_0 \ket  = \lambda |C_0 \ket $, we must have
 ${\bf a} = (000 \, 000 \,000)$  which implies $X({\bf a}) = I$, and 
${\bf b} \cdot {\bf v} = 0$ for every 
${\bf v} \in \Gamma_{6,9}$.  But
since (as noted above)  $\Gamma_{6,9}$ spans  ${\bf Z}_2^9$
this  implies that ${\bf b}$ is orthogonal to all of ${\bf Z}_2^9$,
which implies ${\bf b} = (000 \, 000 \, 000)$ so that 
$Z({\bf b}) = I$ as well.  Thus, since any element of $E$ can be
written as a multiple of $X({\bf a})Z({\bf b})$, the stabilizer
$S$ contains only multiples of the identity.

\section{Conclusion}

Although codes which can correct Pauli exchange errors will be
larger than the minimal 5-qubit codes obtained for single-bit
error correction, this may not be a serious drawback.  
 For implementations of quantum computers which have a grid
structure (e.g., solid state or optical lattices)  it may be natural and
advantageous to use 9-qubit codes  which can be implemented in
$3 \times 3$ blocks. (See, e.g.,  \cite{BCJCZ}.)   
However, codes larger than 9-bits may be impractical
for a variety of reasons.    Hence it is encouraging that both the 
code in section \ref{sect:code9} and the  construction proposed in section 
\ref{sect:newcode} do {\em not} require $n > 9$.

It may be worth
investigating whether or not the codes proposed here 
can be used advantageously in combination with other schemes, 
particularly those \cite{KL2} based on  hierarchical nesting.  
Since the code in sections \ref{sect:code9} and 
\ref{sect:newcode} can already handle some types of multiple
errors, concatenation of one of these 9-bit codes with itself
will contain some redundancy and concatenation with a 5-bit code
may be worth exploring.  Indeed, when exchange correlations are
the prime mechanism for multi-bit errors, the need for repeated
concatenation may be significantly reduced.

Construction of codes of the type proposed in Section \ref{sect:newcode}
remains a significant challenge.  
However, development of such new methods of may be precisely what
is needed to obtain codes powerful enough to correct multi-qubit
errors efficiently, without the large size drawback of codes
based on repeated concatenation.  

\bigskip

\noindent{\bf Acknowledgment}  It is a pleasure to thank Professor
Eric Carlen for a useful comment which started my interest in
exchange interactions,  Professor Chris King for
several helpful discussions and comments on earlier drafts,
 Professor Harriet Pollatsek for additional comments, discussions and
permission to include her observations about the non-additivity of
the 9-bit code presented here, 
Professor Nolan Wallach for communications about  7-bit codes,
and the five anonymous referees of {\em Physical Review Letters}
for their extensive commentary on \cite{Rusk}.


\begin{thebibliography}{~~}

\bibitem{AB} D. Aharanov, M. Ben-Or,
``Fault-Tolerant Quantum Computation With Constant Error Rate''
lanl preprints quant-ph/9611025  and quant-ph/9906129.

\bibitem{BDSW} C.H. Bennett, D.P. DiVincenzo, J.A. Smolin and W.K. Wooters,
``Mixed State Entanglement and Quantum Error Correction''
{\em Phys. Rev. A} {\bf 54}, 3824--3851 (1996)
[lanl preprint quant-ph/9604024].


\bibitem{BCJCZ}  H.J. Briegel, T. Calarco, D. Jaksch, J.I. Cirac,
and P. Zoller 
``Quantum computing with neutral atoms''
lanl preprint quant-ph/9904010.


\bibitem{CRSS} R. Calderbank, E.M. Rains,  P.W. Shor and N.J.A. Sloane,
``Quantum Error Correction and Orthogonal Geometry''
{\em Phys. Rev. Lett.} {\bf 78}, 405--408 (1997) 
[lanl preprint  quant-ph/9605005];  and
``Quantum Error Correction via Codes over GF(4)''
{\em IEEE Trans. Info. Theory} {\bf 44}, 1369--1387 (1998)
[lanl preprint  quant-ph/9608006].

\bibitem{Gott1} D. Gottesman
``Stabilizer Codes and Quantum Error Correction''
PhD thesis, Caltech (1997).
[lanl preprint quant-ph/9705052].

\bibitem{Gott2}  D. Gottesman   
``A Theory of Fault-Tolerant Quantum Computation''
{\em Phys.Rev. A}  {\bf 57}, 127-- (1998)
[lanl preprint quant-ph/9702029].

\bibitem{Gott3} D. Gottesman,
``An Introduction to Quantum Error Correction''
lanl preprint quant-ph/0004072 for AMS short course proceedings,
to appear in {\em Proc. Symp. Appl. Math} (AMS, 2000).

\bibitem{Kt} A. Y. Kitaev
``Fault-tolerant Quantum Computation by Anyons''
lanl preprint quant-ph/9707021.


\bibitem{KL1} E. Knill and R Laflamme, 
``A Theory of Quantum Error-Correcting Codes''
{\em Phys. Rev. A} {\bf 55}, 900-911 (1997).


\bibitem{KL2} E. Knill and R Laflamme, 
``Concatenated Quantum Codes'' lanl preprint quant-ph/9608012.

\bibitem{KLV} E. Knill, R Laflamme, and L. Viola 
``Theory of Quantum Error Correction for General Noise''
{\em Phys. Rev. Lett.}  {\bf 84},  25254--28 (2000)
[lanl preprint quant-ph/9908066].

\bibitem{KLZ}  E. Knill, R Laflamme, W. H. Zurek
 ``Resilient Quantum Computation: Error Models and Thresholds''
{\em Proc. Roy. Soc. A} {\bf 454}, 365--384 (1998).
  [lanl preprint  quant-ph/9702058]

\bibitem{LMPZ} R. Laflamme, C. Miquel, J.P. Paz, W.H. Zurek,
``Perfect Quantum Error Correction Code''
{\em Phys. Rev. Lett.} {\bf 77}, 198--201 (1996).


\bibitem{DFS1} D.A. Lidar, I.L. Chuang,   and K.B. Whaley
``Decoherence Free Subspaces for Quantum Computation''
 {\em Phys. Rev. Lett.}  {\bf 81},  2594--97 (1998)
[lanl preprint quant-ph/9807004].

\bibitem{DFS2} D.A. Lidar, D. Bacon,   and K.B. Whaley
``Concatenating Decoherence Free Subspaces with Quantum Error Correcting Codes''
{\em Phys. Rev. Lett.}  {\bf 82}, 4556--59 (1999)
[lanl preprint quant-ph/9809081].

\bibitem{DFS3} D.A. Lidar, J. Kempe, D. Bacon,   and K.B. Whaley
``Protecting Quantum Information Encoded in Decoherence Free States 
Against Exchange Errors''
{\em Phys. Rev. A} {\bf 61}, 052307 (2000)
[lanl preprint quant-ph/0004064].

\bibitem{DFS4}  D. Bacon, J. Kempe, D.A. Lidar, and K.B. Whaley, 
``Universal Fault-Tolerant Computation on Decoherence-Free Subspaces''
lanl preprint quant-ph/9909058;  and
J. Kempe, D. Bacon,  D.A. Lidar, and K.B. Whaley,
`` Theory of Decoherence-Free Fault-Tolerant Universal Quantum Computation''
lanl preprint quant-ph/0004064.


\bibitem{NC} M.A. Nielsen and I.L. Chuang,
{\em Quantum Computation and Quantum Information}
(Cambridge University Press, in press).

\bibitem{P}  H. Pollatsek, private communication

\bibitem{Pk1} J. Preskill, 
``Reliable Quantum Computers''
{\em Proc. Roy. Soc. A} {\bf 454}, 385--410 (1998)
[lanl preprint quant-ph/9705031], and 
``Fault-Tolerant Quantum Computation ''
lanl preprint quant-ph/9712048

\bibitem{Pk2} J. Preskill, ``Battling Decoherence:
 The Fault Tolerant Quantum Computer''
{\em Physics Today} (6){\bf52}, 24--30 (June, 1999).

\bibitem{RHSS} E.M. Rains, R. H. Hardin,  P.W. Shor and N.J.A. Sloane,
``A nonadditive quantum code''
{\em Phys. Rev. Lett.}  {\bf 79}, 953--954 (1997). 

\bibitem{Rains} E.M. Rains,
``Quantum Codes of Minimum Distance Two''
lanl preprint quant-ph/9704043


\bibitem{RV} V. P. Roychowdhury and F.  Vatan,
``On the Structure of Additive Quantum Codes and the Existence of 
lanl preprint quant-ph/9710031 

\bibitem{Rusk}  M.B. Ruskai,
``Pauli Exchange Errors in Quantum Computation''
{\em Phys. Rev. Lett.} (in press);
lanl preprint quant-ph/9906114


\bibitem{Shor} P. Shor,
``Scheme for Reducing Decoherence in Quantum Computer Memory''
{\em  Phys. Rev. A} {\bf 52}, 2493-2496 (1995). 


\bibitem{St} A.M. Steane,
``Efficient Fault-tolerant Quantum Computing''  
{\em Nature}  {\bf 399}, 124--126 (May 1999).

\bibitem{W} N. Wallach, private communication.


\end{thebibliography}
\end{document}